\documentclass[a4paper,prx,reprint,superscriptaddress,twocolumns,notitlepage,floatfix]{revtex4-2}
\usepackage[utf8]{inputenc}
\usepackage[T1]{fontenc}
\usepackage{amsmath}
\usepackage{amssymb}
\usepackage{braket}
\usepackage{csquotes}

\usepackage{graphicx}
\usepackage{xcolor}
\usepackage[separate-uncertainty = true]{siunitx}
\usepackage{bm}

\usepackage{hyperref}
\usepackage{bookmark}

\def\dd{\text{d}}
\def\ii{\text{i}}
\def\ee{\text{e}}

\hypersetup{%
  breaklinks=true,
  colorlinks=true,
  linkcolor=black,
  filecolor=black,
  urlcolor=black,
  citecolor=black,
  pdfborder=0 0 0,
}

\begin{document}

\title{Internal wave crystals}

\author{Sasan J. Ghaemsaidi}
\affiliation{Massachusetts Institute of Technology, Department of Mechanical Engineering, Cambridge, Massachusetts 02139, U.S.A.}
\author{Michel Fruchart}
\affiliation{Department of Physics and James Franck Institute, University of Chicago, Chicago, Illinois 60637, U.S.A.}
\author{Severine Atis}
\affiliation{Massachusetts Institute of Technology, Department of Mechanical Engineering, Cambridge, Massachusetts 02139, U.S.A.}
\affiliation{Department of Physics and James Franck Institute, University of Chicago, Chicago, Illinois 60637, U.S.A.}
\date{\today}

\begin{abstract}
    Geophysical fluids such as the ocean and atmosphere can be stratified: their density depends on the depth. As a consequence, they can host internal gravity waves that propagate in the bulk of the fluid, far from the surface~\cite{Sutherland2010}.
These waves can transport energy and momentum over large distances, thereby affecting large-scale circulation patterns~\cite{Alford2015}, as well as the transport of heat, sediments, nutrients and pollutants in the ocean~\cite{Sarkar2017}. When the density stratification is not uniform, internal waves can exhibit wave phenomena such as resonances, tunneling, and frequency-dependent transmissions~\cite{Sutherland2004,Ghaemsaidi2016,Sutherland2016}. Spatially periodic density profiles formed by thermohaline staircases are commonly found in stratified fluids ranging from the Arctic Ocean~\cite{Dosser2014} to giant planet interiors~\cite{Belyaev2015,Andre2017}, and can produce extended regions with periodically stratified fluid. Here, we report on the experimental observation of band gaps for internal gravity waves, ranges of frequencies over which the wave propagation is prohibited in the presence of a periodic stratification. 
We show the existence of surface states controlled by boundary conditions and discuss their topological origin.
Our results suggest that energy transport can be profoundly affected by the presence of periodic stratifications in geophysical fluids ranging from Earth's oceans to gas giants.
\end{abstract}

\maketitle


Internal gravity waves are carried in fluids that present density stratification, such as the atmosphere and the oceans~\cite{Sutherland2010}.
They can be generated by the wind, or near the ocean's floor by tidal flow over topography~\cite{Alford2015}. 
Because they can travel long distances before being dissipated, they play an important role in redistributing heat and momentum over large scales, and play a consequent role in the oceanic general circulation and the evolution of climate~\cite{Alford2015,Ferrari2009,Wunsch2004}.
The propagation of internal waves can be strongly affected by a non-uniform stratification.


In the ocean, the interplay between heat diffusion and salt diffusion can lead to double-diffusive instability and produce spatially periodic density profiles called thermohaline staircases over spatially extended regions ~\cite{Radko2013, Dosser2014}. These periodic structures have also been suggested to exist in astrophysical bodies, such as in giant planet interiors~\cite{Belyaev2015,Andre2017}, again as the result of the coexistence of a double-diffusive phenomena~\cite{Radko2013}.
When the size of the steps is small compared with the internal wave wavelength, internal wave propagation can be described by ray theory~\cite{Sutherland2010,Ghaemsaidi2016}.
However, ray theory does not account for wave-like behaviors such as diffraction and tunneling, that can occur when the stratification varies over length scales comparable with the wavelength~\cite{Sutherland2004,Mercier2008,Mathur2010}.
In the Canada Basin, for instance, the period of the thermohaline staircases can be comparable to the wavelength of internal waves~\cite{Cole2014,Dosser2014,Boury2021}.
In this regime, it is expected that a periodic variation of the stratification will strongly affect the propagation of internal waves~\cite{Salusti1978,Malvestuto1979,Sutherland2016,Radko2020}.


\begin{figure*}
\includegraphics[width=18cm]{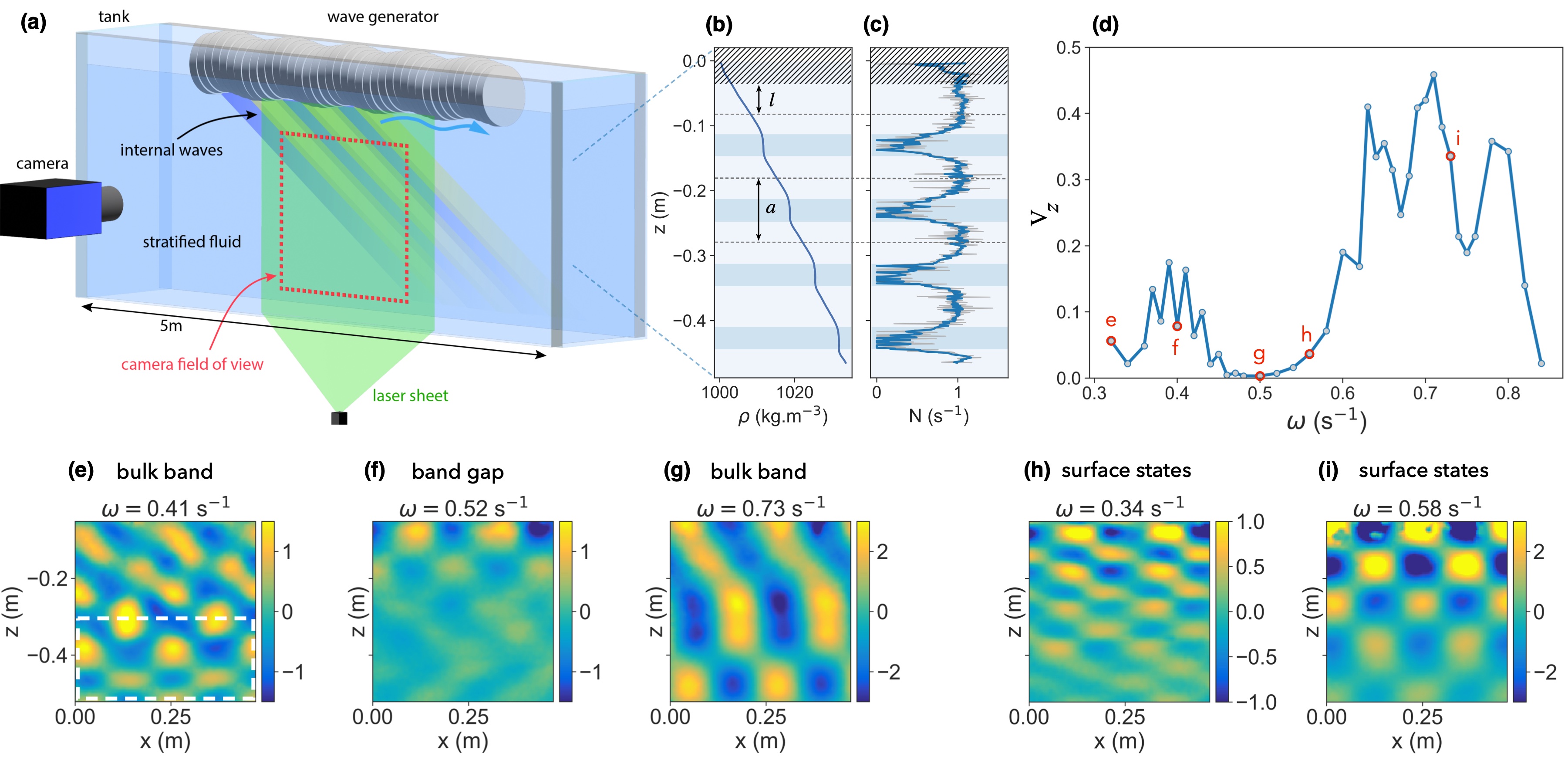}
\caption{\textbf{Experimental observation of a band gap and surface states.} (a) Schematic of the experimental setup with periodic stratification. The submerged wave generator located near the top of the tank creates incident internal waves that propagate downward.
(b) Experimentally measured density profile.
(c) Stratification obtained from panel b. The raw signal is show in gray and a sliding average over a \SI{3}{\milli\meter} window is shown in blue. (d) Transmitted internal wave amplitude versus $\omega$ for $k_x = 27.6$ m$^{-1}$ obtained from the experiment (see Fig.~\ref{fig2} for details).
(e-i) Selected snapshots of the flow velocity field determined from PIV measurements for different values of $\omega$. 
The wave is transmitted to bottom of the tank when the excitation frequency $\omega$ is in the bulk bands (panels e and g), but it quickly decays as a function of $z$ when $\omega$ is in a band gap.
In panels h and i, we show the wave profiles corresponding to surface states. The wave amplitude decays exponentially (see also  Fig.~\ref{fig2}), but the amplitude is appreciably high. This is particularly visible in panel h: artifacts in the PIV due to turbulence are visible near the top.
}
\label{fig1}
\end{figure*}

Motivated by these considerations, we perform laboratory experiments to investigate the dynamics of internal waves in controlled periodic stratifications.
A five meter long wave tank is filled with a mixture of salt water (Fig.~\ref{fig1}a).
The density $\rho_0(z)$ of the fluid as a function of depth is varied with the salt concentration. 
The stratification of the fluid is then characterized by the Brunt-Väisälä frequency, defined as
\begin{equation}\label{N}
	N(z) = \sqrt{-\frac{g}{\rho_0} \, \frac{\dd \rho_0}{\dd z}}
\end{equation}
where $g$ is the acceleration due to gravity, $\rho_0(z)$ is the background density, and the vertical axis $z$ points up.
The stratification is uniform when $N(z)$ is a constant. 
In a nonuniform stratification, the fluid density becomes a non-linear function of $z$ and the buoyancy frequency $N(z)$ can vary with the depth in the fluid.

In the experiment, we prepared a spatially periodic stratification composed of alternating layers with buoyancy frequencies $N_1 \simeq \SI{1.1}{\per\second}$ and $N_2 \simeq 0$, as represented on Fig.~\ref{fig1}a-c (see Methods for experimental details).
A total of \num{4} layers with $N=N_2$ separate \num{3} layers with $N=N_1$.
In addition, a upper and lower layer with $N=N_1$ are present on the top and on the bottom of the tank.
The measured density profile and the resulting buoyancy profile (determined from Eq.~\eqref{N}) are shown in Fig.~\ref{fig1}b-c, and exhibit a spatial period $a = \SI{10(1)}{\centi\meter}$.
We use a submerged linear wave generator~\cite{Gostiaux2006} to generate internal waves with an amplitude $A=\SI{3}{\milli\meter}$ and a dominant horizontal wavenumber $k_x = \SI{27.6}{\per\meter}$ at a given frequency $\omega$.
The resulting waves propagate in the $x-z$ plane, and the corresponding velocity field $\bm{v} = (v_x,0,v_z)$ is measured using particle image velocimetry (PIV). 
Selected snapshots of the recorded wave field $v_z(x,z)$ are displayed in Fig.~\ref{fig1}e-g. 

To estimate how much energy is transmitted through the periodic stratification, we show in Fig.~\ref{fig1}d the average $V_z(\omega)$ of the wave amplitude $|v_z|$ over the region $-z=\num{0.3} \text{--} \SI{0.5}{\meter}$ indicated by the dashed white lines in Fig.~\ref{fig1}e. 
The internal wave transmission exhibits a large drop 
in an extended range of frequencies  $\omega \approx \num{0.45} \text{--} \SI{0.58}{\per\second}$.
For a wave frequency in this range, the transmission is impeded by the stratification, and a rapid decrease of the wave amplitude with depth is seen in Fig.~\ref{fig1}f. 
On the contrary, the wave is fully transmitted through the structure at frequencies outside of this specific range as shown in Figs.~\ref{fig1}e and g.






\begin{figure*}
\includegraphics[width=7in]{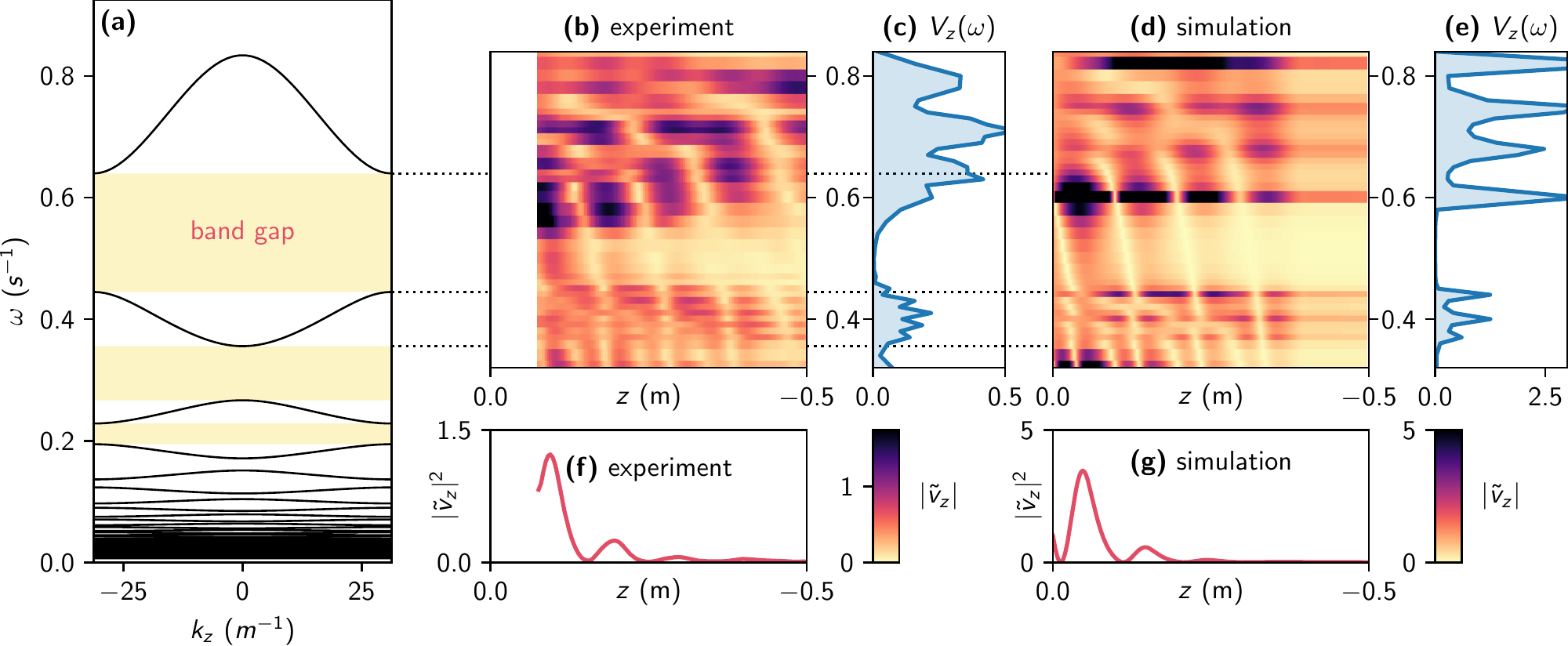}
\caption{\textbf{Band structure and experimental spectrum of the internal wave crystal.}
(a) Internal waves band structure in a periodic stratification determined from the eigenvalues of the operator $\mathcal{L}$ defined by Eq.~\eqref{IW_z} (see Methods). 
(b) Experimental wave amplitude $\tilde{v}_z = v_z/(A \omega)$, in which $A$ and $\omega$ are the amplitude and frequency of motion of the internal wave generator, as a function of depth $z$ and the frequency $\omega$.
(c) To estimate the transmission through the periodic stratification, we compute the average $V_z(\omega)$ of the (normalized) experimental wave amplitude $|\tilde{v}_z|$ over the region between $z\simeq\SI{-0.3}{\meter}$ and $z\simeq\SI{-0.5}{\meter}$, at the bottom of the tank. 
(d) Numerical simulation of the wave amplitude as a function of $z$ and $\omega$.
(e) In this panel, we plot the same quantity as in panel c, this time obtained from numerical simulations.
(f) Exponential decay (accompanied by oscillations) of the surface state measured experimentally, obtained from slicing panel b at $\omega \approx \SI{0.54}{\per\second}$.
(g) Equivalent of panel f in numerical simulations.
}
\label{fig2}
\end{figure*}

This phenomenon can be traced to the existence of a band gap at these frequencies in an infinite version of the crystal realized in our experiment~\cite{Ziman1979}.
To describe the propagation of small-amplitude internal waves, it is convenient to introduce the scalar streamfunction $\psi(x,z,t)$, defined such that 
$v_x = - \partial_z \psi$ and $v_z = \partial_x \psi$.
The propagation of small-amplitude internal waves is then described by the equation (see Ref.~\cite{Sutherland2010})
\begin{equation}
    \label{IW_gen}
    \partial_{t}^2 \nabla^2 \psi + N(z)^2 \partial_{x}^2 \psi = 0.
\end{equation}
For harmonic horizontally periodic solutions of the form $\psi(x,z,t) \propto \phi(z) e^{i(k_x x -\omega t)}$, Eq.~\eqref{IW_gen} reduces
to the ordinary differential equation
\begin{equation}\label{IW_z}
\phi''(z) + k_x^2 \left[ \frac{N^2(z)}{\omega^2} - 1 \right]  \phi(z)=0.
\end{equation}
In a region where $N(z) = N_0$ is constant, the vertical wavenumber $k_z$ of a wave with frequency $\omega$ is given by the dispersion relation $k_z^2(\omega) = k_x^2 \left[ N_0^2 / \omega^2 - 1 \right]$.
In the presence of a periodic stratification $N(z+a) = N(z)$,
the dispersion relation organises into distinct bands $\omega_i(k_z)$ (with $i=1,2,\dots$) separated by band gaps where internal waves cannot propagate.
We use Eq.~\eqref{IW_z} to compute in Fig.~\ref{fig2}a the band structure of an infinite crystal that approximates the experimental stratification profile (see Methods), in which
band gaps are indicated in yellow.
Figure~\ref{fig2}b, displays the experimental wave amplitude as a function of the depth $z$ and the excitation frequency $\omega$. 
Comparing Figs.~\ref{fig2}a and b-c, we find that the low-transmission regions in the experimental spectrum of Figs.~\ref{fig2}b qualitatively match the band gaps predicted analytically. 
However, the experimental band gap appears to be smaller than the one predicted analytically. This observation persists in direct numerical simulations of internal wave propagation shown in Figs.~\ref{fig2}d-e (see Methods). 
Taking a closer look at Fig.~\ref{fig2}b and d, we note that a large wave amplitude is present inside the predicted bulk band gap around $\omega \sim \SI{0.6}{\per\second}$ suggesting the existence of a resonance at this frequency.
The exponential decay with depth of the wave amplitude in Fig.~\ref{fig2}f-g shows that this resonance is localized near the upper boundary of the system.

\begin{figure*}
\includegraphics[width=7in]{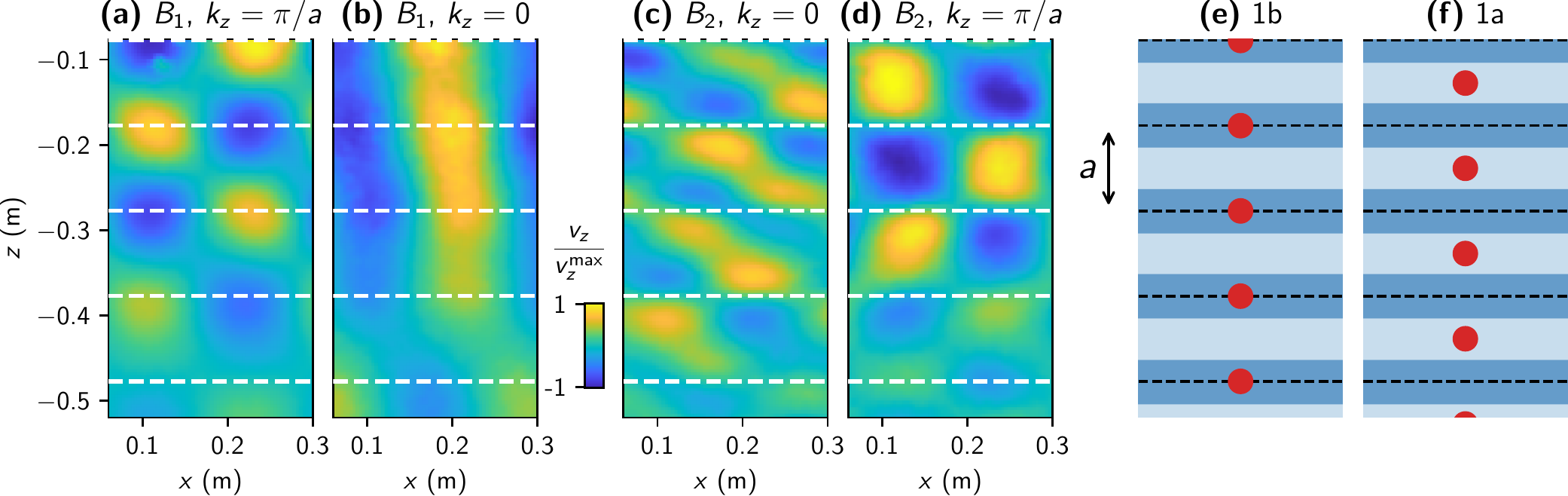}
\caption{\textbf{Topological invariant in the internal wave crystal.}
To compute the Zak phase of the one-dimensional internal wave crystal, we measured experimentally the Bloch functions $v_z(x, z ; k_z)$ of the two highest bands at momentum $k_z = 0$ and $k_z = \pi/a$ (normalized by their maximal values $v_z^{\text{max}}$) in panels a-d.
The boundaries of the unit cells are marked by white dashed lines.
Panels a-b (c-d) describe the highest (second highest) band.
In both cases, the Bloch function $v_z(x, z ; k_z)$ at $k_z = 0$ is antisymmetric with respect to the center of the unit cell, while the Bloch function at $k_z=\pi/a$ is symmetric. Hence, the Zak phases of both bands are $\pi$.
This means that the Wannier centers of the bands are located at the boundary of the unit cell (called the Wyckoff position 1b), as represented in panel e.
In panel f, we represent the other possible situation (not occurring in the experiment) in which the Wannier centers are located in the center of the unit cell (called the Wyckoff position 1a).
}
\label{fig3}
\end{figure*}

The surface states shown in Fig.~\ref{fig2} are reminiscent of surface states found in topological insulators~\cite{Hasan2010,Bradlyn2017}, in contexts ranging from electrons in solids~\cite{Hasan2010} to optics~\cite{Ozawa2019,Lu2014} and geophysical fluids~\cite{Delplace2017,Perrot2019}.
To understand this connection, notice first that the periodic stratification $N(z)$ is approximately invariant under spatial inversion: $N(z_0 + z) = N(z_0-z)$, where $z_0$ is at the center of a unit cell (see Fig.~\ref{fig1}c).
In this case, one can define a topological invariant called the Zak phase $\varphi_{\text{Zak}}$ (there is one Zak phase per band). It is constrained to take values $0$ or $\pi$, which label the two topologically distinct states.
Qualitatively, the Zak phase gives the average position of the wave in the unit cell (called the Wannier center of the band, see Refs.~\cite{Zak1989,Bradlyn2017,Cano2021,Wieder2021} for details).
In a system with inversion symmetry, the Wannier center can only be located at the center ($\varphi_{\text{Zak}}=0$) or at the edge ($\varphi_{\text{Zak}}=\pi$) of the unit cell.
Consider a finite system terminated at the edge of a unit cell (black dashed lines in Fig.~\ref{fig3}e-f). 
A surface state occurs when a Wannier center is exposed at the boundary, like in Fig.~\ref{fig3}e. In contrast, no surface state occurs when the Wannier center is in the middle of the unit cell, like in Fig.~\ref{fig3}f.

The Zak phase can be directly measured experimentally from the wave fields using its expression~\cite{Zak1989}
\begin{equation}
	\label{zak_phase}
	\frac{\varphi_{\text{Zak}}}{\pi} = \frac{\xi_0 - \xi_\pi}{2} \, \text{mod $2$}.
\end{equation}
Here, the quantities $\xi_{k_z}$ take the values $\pm 1$ depending on whether the Bloch wave $\phi(k_z)$ at $k_z=0, \pi/a$ are symmetric or antisymmetric by reflection about the center of the unit cells. 
Figures~\ref{fig3}a-d are showing snapshots of the experimental Bloch waves in which the boundaries of the unit cells are indicated by white dashed lines.
Starting with the highest frequency band $B_1$, we find from Fig.~\ref{fig3}a-b that $\xi_0 = -1$ and $\xi_\pi = 1$, leading to $\varphi_{\text{Zak}} = \pi$. Similarly, we find $\varphi_{\text{Zak}} = \pi$ for the second highest frequency band $B_2$ (Fig.~\ref{fig3}c-d). 
In this configuration, we would expect to observe a surface state if the experimental system was terminated at the edge of a unit cell (i.e., at one of the white dashed lines in Fig.~\ref{fig3}c-d).

\begin{figure}
\includegraphics[width=3.4in]{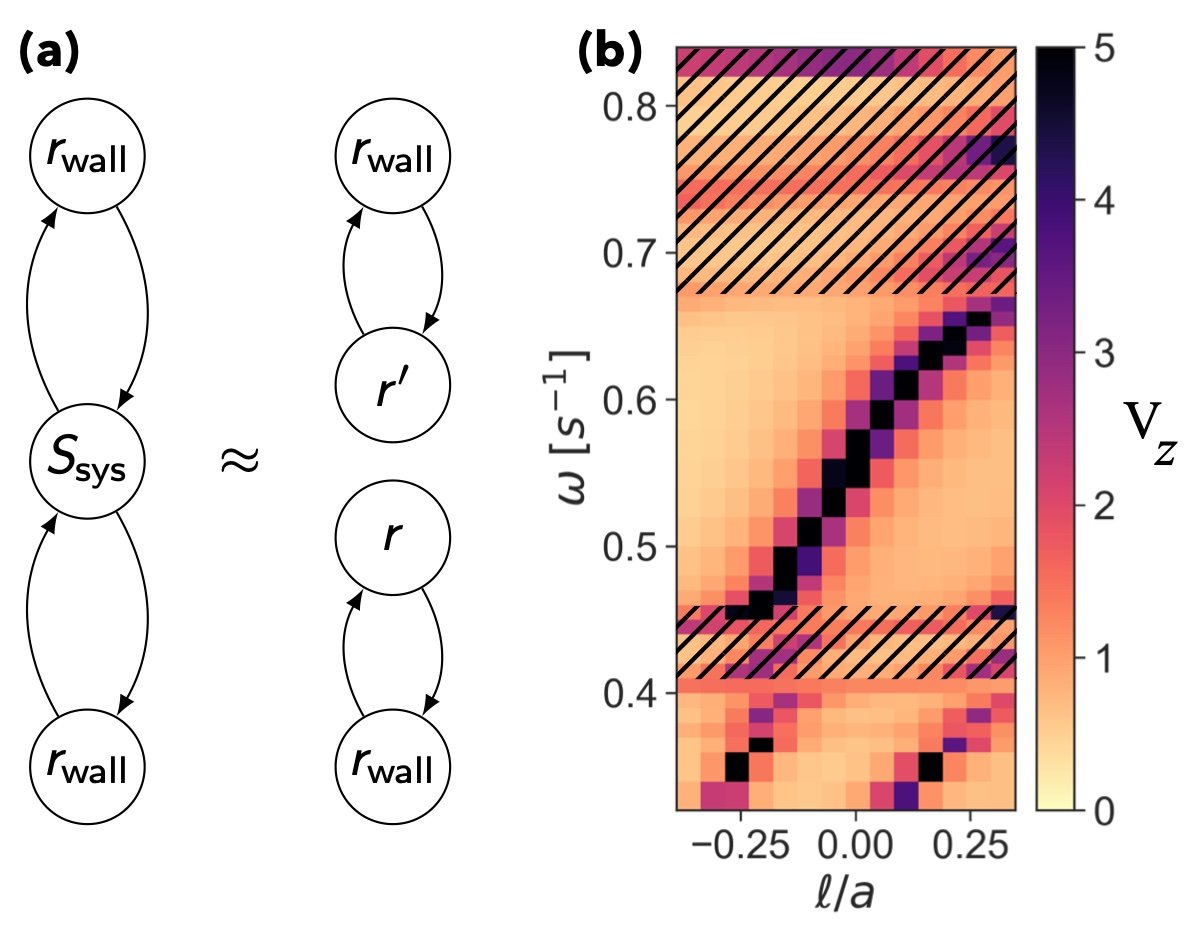}
\caption{\textbf{Effect on the upper layer depth on surface states.}
(a) Scattering matrix model of the finite stratified structure. From the point of view of the waves, the finite system composed of the internal wave crystal, plus walls (at the bottom of the tank, and at the wave generator). It can be modelled by a graph of scattering matrices: $S_{\text{sys}}$ represents the propagation in the periodic stratification (plus the upper and lower layers), while the reflection coefficients ($1 \times 1$ scattering matrices) $r_{\text{wall}}$ represent the effect of the hard walls.
When the system is gapped, it acts as a frequency-dependent mirror from the point of view of the outside. Hence, the top and bottom of the systems are decoupled and $S_{\text{sys}}$ can be replaced by frequency-dependent reflection coefficients $r(\omega)$ and $r'(\omega)$ on each end.
(b) We show a plot of the transmission through the system obtained from numerical simulations (in the same way as Fig.~\ref{fig2}e) as a function of the depth of the last layer $\ell/a$.
The bulk bands are marked by dashed patterns.
The frequency of the surface state changes from $\omega \simeq \SI{0.45}{\per\second}$ (bottom of the gap) to $\omega \simeq \SI{0.68}{\per\second}$ (top of the gap) when $\ell/a$ increases, as the distance of Wannier center to the surface changes.
We note that there are also surface states in the second gap (below \SI{0.4}{\per\second}).
}
\label{fig4}
\end{figure}

However, in the experiment, the wave generator is not located at the edge of a unit cell: there is an additional layer of size $\ell$ with stratification $N_1$ (see Fig.~\ref{fig1}). Hence, the Wannier centers are nor at the center, neither at the boundary of the unit cell. Although it is tempting to expect that a surface state will still exist in these intermediate situations, this cannot be determined from considerations of topology alone.
Instead, the existence of surface states can be phrased as a general condition of constructive interference~\cite{Kaliteevski2007,Xiao2014,Levy2017}.
In our experiments, the upper layer $\ell$ can be viewed as a cavity bounded by two mirrors with reflection coefficients $r(\omega)$ and $r_{\text{wall}}$, as represented in Fig.~\ref{fig4}a (see also Methods). The reflection coefficient $r(\omega) = \ee^{\ii \varphi(\omega)}$ represents the periodic stratification, that acts as a frequency-dependent mirror in the gap. The boundary condition $\psi = 0$ at a wall (such as the wave generator) fully reflects the wave with a  reflection coefficient $r_{\text{wall}} = - 1$. 
Resonant modes then occur when a wave interferes constructively with itself over a round-trip, i.e. when the total phase $\varphi_{\text{tot}} = \varphi(\omega) + 2 k_1(\omega) \ell + \pi $ is an integer multiple of $2\pi$.

To monitor the evolution of surface states with the depth $\ell$ of the upper layer, we artificially vary $\ell$ in numerical simulations. 
The wave amplitude near the surface is shown in Fig.~\ref{fig4}b as a function of $\omega$ and $\ell$.
We observe that the frequency of the surface states fully crosses the band gap as $\ell/a$ is increased.
In the Methods, we analyse a simplified model of the experiment, in which these features can be reproduced.
Remarkably, the edge states always flow through the band gap in the same direction as we vary the parameter $\ell/a$, a bit like in a Thouless pump~\cite{Thouless1983}.
Although we focus here on the surfaces states occurring near the top surface, because they can easily be accessed experimentally, we emphasize that edge modes can in general also exist at the bottom of the periodic stratification, as well as at the interface between two different periodic stratifications, in the absence of any hard boundary.



Our work demonstrates the existence of internal waves featuring band gaps described by band theory as well as surface states described by topological band theory in periodic stratifications.

A band gap could completely prevent the propagation of internal waves at certain frequencies in regions with periodic stratification. 
The presence of an interface state would allow horizontal propagation, but only in a restricted vertical region near the interface and in a narrow range of frequencies, in a way similar to internal wave ducting~\cite{Sutherland2010}.
These effects, if they indeed occur outside of the lab, could have noticeable consequences in regions such as the Arctic ocean, where staircase stratifications with the appropriate length scale exist~\cite{Cole2014,Dosser2014}.
Similar phenomena are expected when the periodicity of the stratification is imperfect. Irregular thermohaline staircases can in principle exhibit Anderson localization, another phenomenon that hinders the transmission of waves~\cite{Evers2008}.



\section*{Methods}

\subsection{Experimental setup}

We first investigate the effect of periodically stratified fluid on internal wave propagation with a laboratory experiment in a $H \times L \times W = 0.54 \times 5.46 \times \SI{0.55}{\cubic\meter}$ wave tank filled with salt-stratified water. To avoid reflections from the sides of the tank to interact with the waves generated in the working section, parabolic reflection barriers are positioned at both ends of the tank and deviate the waves to the rear section of a partition running along the length of the tank \cite{Ghaemsaidi2016}, as shown in the schematic of Fig. \ref{fig1}(a). We use a traditional double-bucket method to fill the tank and create a non-uniform stratification composed of 5 layers of linearly varying density giving $N_1 \simeq 1.1$ s$^{-1}$, separated by 4 mixed layers of constant density where $N_2 \simeq 0$, see Fig.~\ref{fig1}(b).
The density profile $\rho_0(z)$ shown in Fig.~\ref{fig1}(c) is measured using a calibrated Precision Measurements Engineering probe attached to a vertically held Parker linear traverse. The resulting buoyancy profile determined from Eq.\eqref{N} displayed in Fig.~\ref{fig1}(c) exhibits a three layer structure with a periodicity $a = 10 \pm \SI{1}{\centi\meter}$ surrounded by upper and lower layers with identical $N$ on top and bottom of the tank respectively. Note that the last $\SI{5.6}{\centi\meter}$ of the density profile is not shown since our probe doesn't reach the bottom of the tank.
We generate an internal wave field with a dominant characteristic horizontal wavenumber $k_x = \SI{27.6}{\per\meter}$ using a submerged linear wave generator with a sinusoidal profile with an amplitude $A=\SI{3}{\milli\meter}$ placed at the top of the tank.
The waves are horizontally forced from left to right and propagate downward through the periodic stratification in a range of frequency $0.32<\omega<\SI{0.84}{\per\second}$ set by the generator.
The flow velocity field across a $50 \times \SI{50} {\square\centi\meter}$ domain is then measured with particle image velocimetry (PIV), allowing a direct visualization of the internal wave spatial structure inside the periodic stratification as a function of the wave frequency $\omega$. Selected snapshots of recorded wave fields are displayed in Fig. \ref{fig1}(c) for 4 different frequencies.

\subsection{Numerical simulations}

We impose a downward propagating internal wave in the upper boundary and use the experimentally measured stratification profile to solve for the transmitted wave amplitude at the lower boundary. The wave at the top boundary is imposed to be equal to the horizontal wave produced by the generator $\Psi(x,t) = \Psi_0 \exp(ik_xx)$ and evaluate Eq. \ref{IW_gen} as a boundary problem. The subsequent set of equations are then numerically solved using MATLAB \textit{bvp4c} function. The resulting transmitted wave amplitude is shown in \ref{fig1}(d) and is in good agreement with the experimental band structure. Figure \ref{fig2}(c) displays the resulting wave field amplitude as a function of $z$ and $\omega$, and exhibits two distinct maxima localized near the upper boundary located near similar frequencies, $\omega \approx 0.33$ and $\SI{0.6}{\per\second}$, with the experimental surface states.
We now vary the thickness $\ell$ of the top layer in the numerical model, and re-iterate the same frequency sweep shown in Fig. \ref{fig2}(c); see Supplemental material for the corresponding wave fields. Changing $\ell$, we expect the resonance conditions to shift accordingly and Fig. \ref{fig4}(c) shows the resulting amplitude of the internal wave in the upper layer as we change $\ell$. While the band structure appears unchanged, the two edge modes frequency location seems to shift as $\ell$ is increased. Remarkably, both of the band gap crossings are happening in the same direction, while additional edge modes continue to appear as we vary the parameter $\ell/a$.

\subsection{Computation of the band structures}
\label{computation_bs_spectral}

In this section, we describe the computation of the band structure of a model of the experiment based on the experimentally measured density profile. As the stratification is not exactly piecewise constant, we found more convenient to directly compute the band structure by diagonalizing a linear operator, rather than using the method based on transfer matrices used to analyse the simplified model of section \ref{app_model_interface_states}.

We start from the one-dimensional differential equation~\eqref{IW_z}, and assume that the stratification $N(z) = N(z+a)$ is spatially periodic with period $a$. 
The propagation of internal waves in an infinite system with such a periodic stratification is described by Bloch theory~\cite{Ziman1979,Salusti1978}. 
The main result of this theory is that the streamfunction takes the form of a plane wave modulated by a function with the same periodicity as the unit cell (called a Bloch wave)
\begin{equation}
\phi(z) = \ee^{\ii k_z z} u(z)	
\end{equation}
in which $u(z) = u(z+a)$.
Here, the quasi-momentum (or wavevector) $k_z$ lives in the Brillouin zone $[-\pi/a, \pi/a]$.
The propagation of waves in finite crystals is usually well-approximated by the infinite crystal (except from interface states that are discussed in Methods Sec.~\ref{app_model_interface_states}).
Equation~\eqref{IW_z} then takes the form of an eigenvalue problem
\begin{equation}
	\mathcal{L}(k_x,k_z) \ket{u} = \lambda \ket{u}
\end{equation}
in which $\mathcal{L}(k_x,k_z)$ is a differential operator acting on periodic functions $[0,a] \to \mathbb{R}$ defined by
\begin{equation}
	\label{linear_operator_waves}
	\mathcal{L}(k_x,k_z) \ket{u} = \frac{1}{N^2(z)} \left[ k_x^2 - (\partial_z + \ii k_z)^2 \right] \ket{u}
\end{equation}
and where the eigenvalue is $\lambda = k_x^2/\omega^2$.
Diagonalizing $\mathcal{L}(k_x,k_z)$ gives the dispersion relations $\omega_i(k_x, k_z)$ where the integer $i$ labels the band.
To do so, we discretize Eq.~\eqref{linear_operator_waves} using the open-source pseudospectral solver Dedalus~\cite{Burns2020}. The one-dimensional domain $[0,a]$ is discretized with $n_{\text{C}}=2^8$ Chebyshev modes under periodic boundary conditions. The resulting matrix is diagonalized.
Because of the nature of internal waves, the band structure is bounded from above (there is a maximum frequency). 
Low-frequency bands correspond to modes that oscillate fast at the scale of a unit cell. 
Hence, the finite number $n$ of Chebyshev modes allows us to resolve only so many bands, starting from the highest-frequency ones (this is in contrast with the spectral theory of the Laplacian and derived problems, in which the lowest-frequency bands correspond to slow oscillations in the unit cell and would be resolved first).
As seen in Fig.~\ref{fig2}a, the low-frequency bands accumulate near $\omega = 0$ to form a quasi-continuum.

We computed the band structure of the infinite periodic stratification that would arise from an infinite repetition of one unit cell of the stratification measured in the experimental system (Fig.~\ref{fig2}a). 
The stratification $N(z)$ is obtained by fitting the experimental density profile and computing the derivative of the fit analytically, while the other parameters $a$ and $k_x$ are measured directly.

\subsection{Simplified model of interface states}
\label{app_model_interface_states}

In this section, we provide a simplified model for stratifications with interface states as well as general arguments based on scattering matrices for their existence adapted from Refs.~\cite{Kaliteevski2007,Lawrence2009,Lawrence2010,Xiao2014,Levy2017,Lee1981,Dwivedi2016}.



\subsubsection{Scattering matrices}

\begin{figure}
\includegraphics{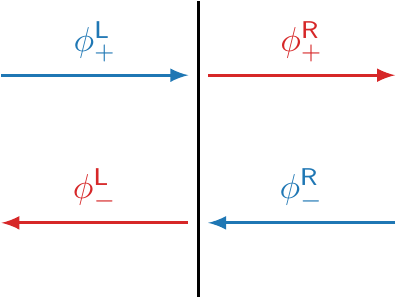}
\caption{\label{scattering_notations} \textbf{Scattering at an interface.}
The interface is drawn in black.
Incoming amplitudes are drawn in blue while outgoing amplitudes are drawn in red.
The amplitudes of right-moving (left-moving) waves are labelled with a $+$ ($-$).
}
\end{figure}

We want to keep track of the phases over propagation but also decompose the function into left/right-moving fields, so to get the amplitude of the field at $z_0$, we decompose the field as
\begin{equation}
    \phi(z) = \phi_{+} \ee^{\ii k(\omega) (z-z_0)} + \phi_{-} \ee^{-\ii k(\omega) (z-z_0)}
\end{equation}

Consider an interface between two media with dispersions $k_{\text{L/R}}(\omega)$, where L/R correspond to the left/right of the interface. (To match with usual notations for scattering matrices, we consider that the propagation is horizontal to define the notations. To get back to physical directions, one can substitute left/right with bottom/top, respectively.)

A plane wave with frequency $\omega$ on the left/right of the interface takes the form
\begin{equation}
    \phi^\text{L/R}(z) = \phi_{+}^\text{L/R} \ee^{\ii k_{\text{L/R}}(\omega) z} + \phi_{-}^\text{L/R} \ee^{-\ii k_{\text{L/R}}(\omega) z}
\end{equation}
(the full time-dependent wave also has a phase $\ee^{- \ii \omega t}$, already factored out of $\phi$).

The scattering matrix $S$ relates incoming waves (incident on the interface) and outgoing waves (scattered by the interface) by
\begin{equation}
	\begin{pmatrix}
	\phi_{+}^\text{R}
	\\
	\phi_{-}^\text{L}
	\end{pmatrix}
	= 
	S
	\begin{pmatrix}
	\phi_{-}^\text{R}
	\\
	\phi_{+}^\text{L}
	\end{pmatrix}.
\end{equation}
The notations are summarized in Fig.~\ref{scattering_notations}.
We refer the reader to Refs.~\cite{Markos2008,Schomerus2018} for more details.

The scattering matrix can be expressed as
\begin{equation}
	S = \begin{pmatrix}
	r & t' \\
	t & r'
	\end{pmatrix}
\end{equation}
in which $r$ and $t$ ($r'$ and $t'$) describe the reflection and transmission of waves arriving from the left (from the right).
In general, $r$, $r'$, $t$, $t'$ can be matrices (here we only consider one channel, so they are just numbers).

Equivalently, the transfer matrix
\begin{equation}
	T = \begin{pmatrix}
		(t^\dagger)^{-1} & r' (t')^{-1} \\
		(r')^\dagger (t^\dagger)^{-1} & (t')^{-1}
	\end{pmatrix}	
\end{equation}
(in which $\dagger$ is the conjugate-transpose)
relates the fields on the left to the fields on the right by
\begin{equation}
	\begin{pmatrix}
	\phi_{+}^\text{R}
	\\
	\phi_{-}^\text{R}
	\end{pmatrix}
	= 
	T
	\begin{pmatrix}
	\phi_{+}^\text{L}
	\\
	\phi_{-}^\text{L}
	\end{pmatrix}.
\end{equation}

The transfer matrix of two adjacent regions $(1)$ and $(2)$ with transfer matrices $T_1$ and $T_2$ is $T_2 T_1$ (matrix multiplication).
The composition law of scattering matrices can be deduced from that of scattering matrices \cite[Eq.~(10.35)]{Schomerus2018}.

\subsubsection{Model and transfer matrix of the unit cell}

\begin{figure}
\includegraphics{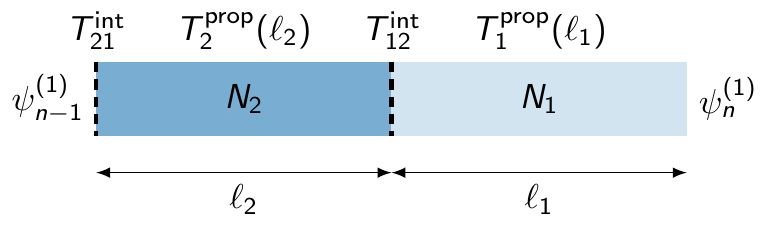}
\caption{\label{transfer_matrix_unit_cell} \textbf{Unit cell of the simplified model.}
}
\end{figure}

Each unit cell of the crystal is composed of two layers of depth $a/2$ with uniform stratifications $N_1$ and $N_2$ (Fig.~\ref{setup_surface_modes_scattering}).
Within each layer, the dispersion relation of internal waves is
\begin{equation}
	k_{z, a}(\omega) = \sqrt{k_x^2 \left( \frac{N_a^2}{\omega^2} - 1 \right)}
\end{equation}
in which $a=1,2$.
To describe the system, we use the formalism of transfer matrices: the transfer matrix $T_{\text{uc}}$ of a single unit cell maps the fields $\psi_{n-1}^{(1)} = (\phi_{+,n-1}^{(2)}, \phi_{-}^{(1)})$ in the first layer of the $(n-1)$th unit cell to the fields $\psi_{n}^{(1)}$ in the first layer of the $n$th unit cell by
\begin{equation}
	\label{def_Tuc}
	\psi_{n}^{(1)} = T_{\text{uc}} \psi_{n-1}^{(1)}
\end{equation}
It is decomposed into (see Fig.~\ref{transfer_matrix_unit_cell})
\begin{equation}
	T_{\text{uc}} = T_{1}^\text{prop}(\ell_1) \, T_{1 2}^\text{int} \, T_{2}^\text{prop}(\ell_2) \, T_{2 1}^\text{int}
\end{equation}
in which 
\begin{equation}
	T_{a}^\text{prop}(\ell) = \begin{pmatrix}
	\ee^{\ii k_a(\omega) \ell} & 0 \\
	0 & \ee^{-\ii k_a(\omega) \ell} \\
	\end{pmatrix}
\end{equation}
describes the propagation within a layer of size $\ell$ with dispersion $k_a(\omega)$ and $T_{a b}^\text{int}$ the interface between the layers $a$ and $b$.
We assume that the density $\rho$ is continuous at each interface, so that the streamfunction $\psi$ and its derivative $\psi'$ are continuous \cite{Sutherland2010}. Hence, we find that at an interface between layers with dispersions $k_L(\omega)$ on the left of the interface and $k_R(\omega)$ on the right, 
\begin{align}
	\phi_{+}^\text{L} + \phi_{-}^\text{L} &= \phi_{+}^\text{R} + \phi_{-}^\text{R} \\
	\frac{1}{k_L} \left[ \phi_{+}^\text{L} - \phi_{-}^\text{L} \right] &= \frac{1}{k_R} \left[ \phi_{+}^\text{R} - \phi_{-}^\text{R} \right].
\end{align}
These relations can be written in the form
\begin{equation}
	\begin{pmatrix} \phi_{+}^\text{R} \\ \phi_{-}^\text{R} \end{pmatrix} = T_{R L}^\text{int} \begin{pmatrix} \phi_{+}^\text{L} \\ \phi_{-}^\text{L} \end{pmatrix}
\end{equation}
in which we have defined the transfer matrix
\begin{equation}
	T_{R L}^\text{int} = \frac{1}{2 k_\text{L}} \, 
	\begin{pmatrix}
		k_\text{L}+k_\text{R} & k_\text{L}-k_\text{R} \\
 		k_\text{L}-k_\text{R} & k_\text{L}+k_\text{R}
	\end{pmatrix}
\end{equation}

\subsubsection{Bulk band structure}

\begin{figure}
\includegraphics{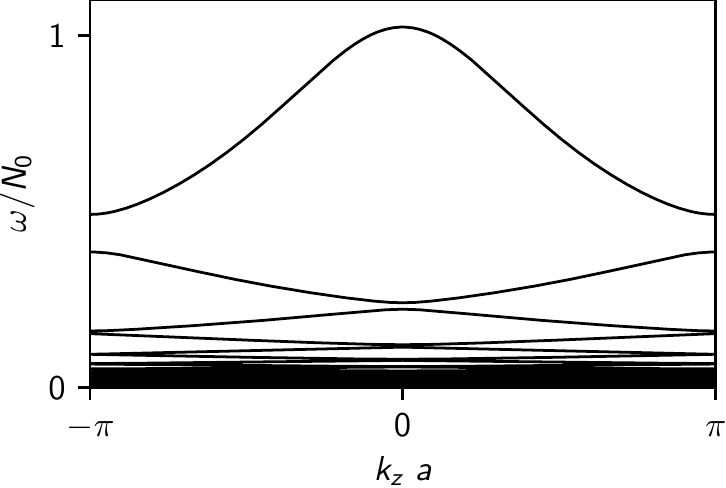}
\caption{\label{band_structure_simplified_model} \textbf{Band structure of the simplified model.}
The band structure is obtained by diagonalizing $T_{\text{uc}}(\omega)$. The eigenvalues satisfying $|\lambda| = 1$ correspond to Bloch modes with a wavevector $k_z$ such that $\lambda = \ee^{\ii k_z \, a}$. 
Frequencies $\omega$ for which no eigenvalue satisfies $|\lambda| = 1$ correspond to band gaps.
We have set $N_2/N_0 = \num{1.2}$, $N_1/N_0 = \num{0.8}$, $k_x/a = \num{3.0}$.
}
\end{figure}

The properties of the infinite crystal (and of the bulk of a large enough finite crystal) are described by Bloch theory.
Bloch waves satisfy
\begin{equation}
	\psi_{n}^{(1)} = \ee^{\ii k_z \, a} \psi_{n-1}^{(1)}
\end{equation}
in which $a = \ell_1 + \ell_2$ is the size of the unit cell, and $k_z$ the quasi-momentum in the direction of propagation $z$.
Comparing with Eq.~\eqref{def_Tuc}, we identify the Bloch phase factors $\ee^{\ii k_z \, a}$ with the eigenvalues $\lambda$ of the transfer matrix $T_{\text{uc}}$ that satisfy $|\lambda| = 1$. The corresponding eigenvectors $\ket{\Psi}$ satisfying
\begin{equation}
	T_{\text{uc}} \ket{\Psi} = \lambda \ket{\Psi}
\end{equation}
are the Bloch eigenvectors.
We can therefore obtain the dispersion relation $k_z(\omega)$ [and by inverting this relation, $\omega(k_z)$] by diagonalizing $T_{\text{uc}}$ and selecting the relevant eigenvalues. The band structure obtained in this way is plotted in Fig.~\ref{band_structure_simplified_model}. As illustrated in Fig.\ref{band_structure_simplified_model}, the band structure features band gaps, a internal waves with frequencies in the band gap cannot propagate through the crystal.

\subsubsection{Surface modes}

\begin{figure}
\includegraphics{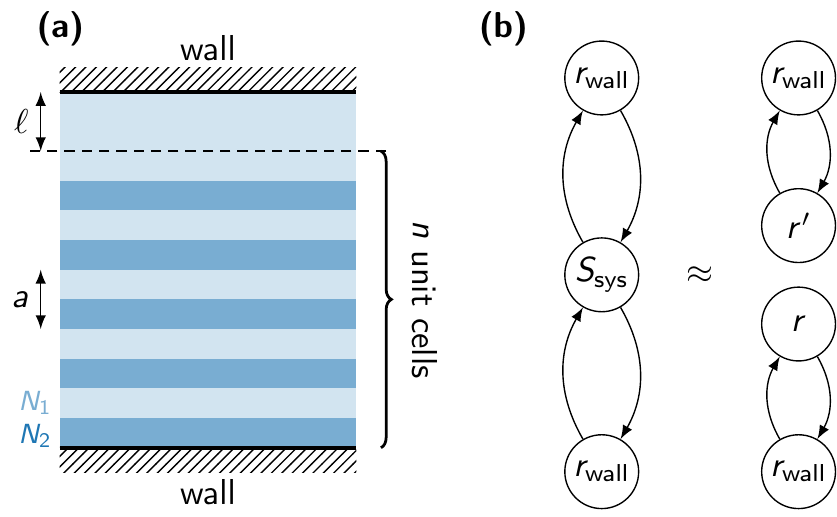}
\caption{\label{setup_surface_modes_scattering} \textbf{Scattering matrix model of the finite stratified structure.}
(a) Schematic representation of the model. A unit cell of length $a$ composed of two layers with stratifications $N_1$ and $N_2$ is repeated.
In the infinite crystal, the unit cell is repeated an infinite number of times.
In a finite system, it is only repeated a certain number $n$ of times.
Here, we assume that the system is terminated by hard walls (like in the experiment), and allow the possibility that the last layer (at the top) has a thickness $\ell_2 + \ell$ instead of $\ell_2$ (again, like in the experiment).
(b) From the point of view of internal waves, this finite system can be modelled by a graph of scattering matrices: $S_{\text{sys}}$ represents the propagation in the periodic stratification (plus the last layer), while the reflection coefficients ($1 \times 1$ scattering matrices) $r_{\text{wall}}$ represent the effect of the hard walls.
When the system is gapped, the top and bottom of the systems are decoupled and  $S_{\text{sys}}$ can be replaced by a reflection coefficient on each end ($r$ and $r'$).
}
\end{figure}

Now, we consider a finite structure composed of $n$ unit cells, but in which the last (upper) layer has a depth $a/2 + \ell$ instead of $a/2$ (see Fig.~\ref{setup_surface_modes_scattering}a).
At the top and bottom of the structure, a wall imposes the boundary condition $\psi = 0$.
We decompose this structure into (i) the lower wall, (ii) the crystal, (iii) the additional layer of depth $\ell$ and (iv) the upper wall; and focus on what happens at the top.

When $\omega$ is in the bulk band gap of the stratified structure, there is no transmission of internal waves through the crystal. 
Hence, the scattering matrix of the crystal is of the form
\begin{equation}
	S_{\text{crystal}} = \begin{pmatrix}
	r_{\text{crystal}} & 0 \\
	0 & r'_{\text{crystal}}
	\end{pmatrix}
\end{equation}
(as $t = t' = 0$; when the crystal is finite, the transmission coefficients do not strictly vanish and taking them to zero is an approximation).
If there is no absorption, the reflection coefficients are simply phases $r = \ee^{\ii \varphi_0}$ and $r' = \ee^{\ii \varphi_0'}$.
In the additional layer of thickness $\ell$, internal waves freely propagate: this is described by the transfer matrix
$T_{\text{layer}} = \operatorname{diag}(\ee^{\ii \varphi}, \ee^{-\ii \varphi})$ or equivalently by the scattering matrix
\begin{equation}
	S_{\text{layer}} = \begin{pmatrix}
	0 & \ee^{\ii \varphi} \\
	\ee^{\ii \varphi} & 0
	\end{pmatrix}
\end{equation}
in which $\varphi = k_2(\omega) \ell$.
Hence, the scattering matrix for the full structure is
\begin{equation}
	S_{\text{sys}} \equiv \begin{pmatrix}
	r & 0 \\
	0 & r'
	\end{pmatrix}  = \begin{pmatrix}
	r_{\text{crystal}} & 0 \\
	0 & \ee^{2 \ii \varphi} r'_{\text{crystal}}
	\end{pmatrix}
\end{equation}
so that the reflection phase of waves arriving from the right (physically, from the top) is $\varphi_0' + 2 \varphi$.

\begin{figure}
\includegraphics[width=2.8in]{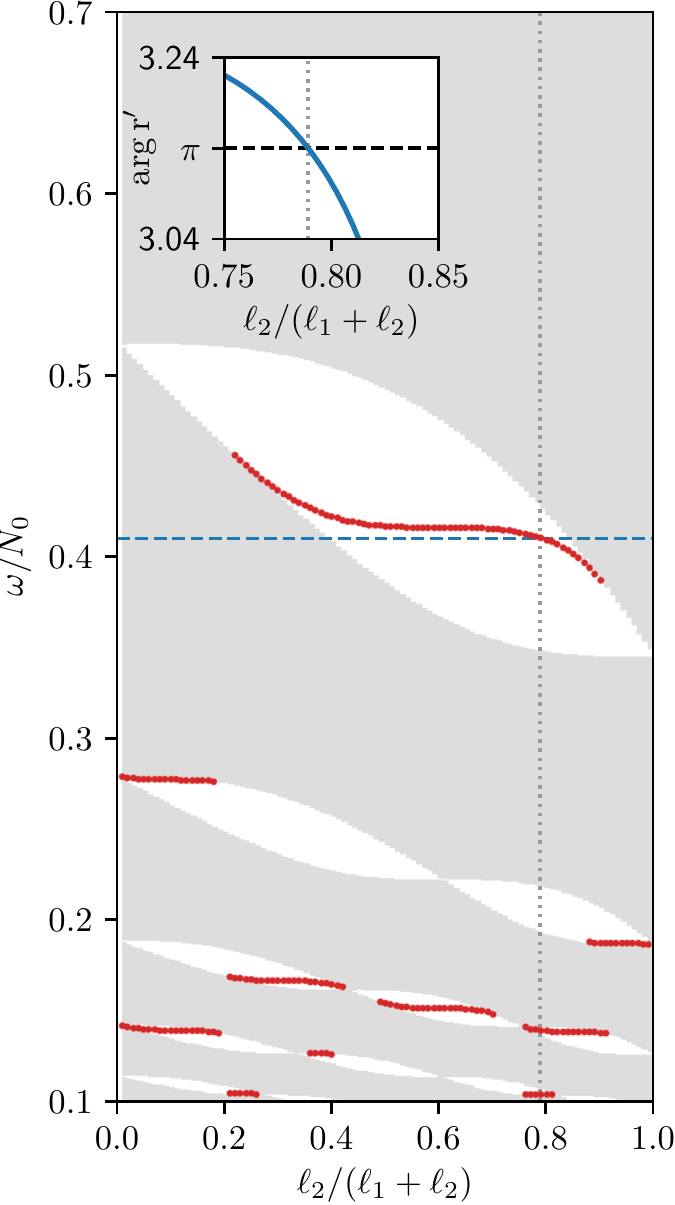}
\caption{\label{surface_states_simplified_model_uc_ratio} \textbf{Surface states of the simplified model.}
Surface states of the simplified model with fixed $\ell = \num{0.2}$ as a function of the relative thickness $\ell_2/[\ell_1+\ell_2]$ of the two layers in a unit cell, computed using the transfer matrix method~\cite{Lee1981,Dwivedi2016}.
The bands are marked as grey regions, and the surface states as red curves.
We have only drawn the surface states (in red) at a single wall, where the boundary condition $\psi = 0$ is imposed.
Inset: reflection phase $\arg r'$ of a finite crystal with $n=\num{20}$ unit cells for $\omega/N_0 = \num{0.41}$ (corresponding to the blue dashed line in the main figure).
Surface states occur when the reflection phase crosses $\pi$. The corresponding value of $\ell_2/[\ell_1+\ell_2]$ is marked by a gray dotted line both in the inset and in the main figure.
We have set $N_1/N_0 = \num{1.2}$, $N_2/N_0 = \num{0.8}$, $k_x/a = \num{3.0}$.
}
\end{figure}

As there is no transmission of internal waves through the stratification described by $S_{\text{sys}}(\omega)$ at frequencies $\omega$ in the bulk band gap of the crystal, we can analyse the top and the bottom independently (Fig.~\ref{setup_surface_modes_scattering}b). 
For concreteness, let us focus on the top (a similar analysis applies to the bottom).
The reflection phase due to the upper wall is $\pi$~\footnote{At the wall, located e.g. at $z=0$, we impose $\psi = 0$. 
Hence, $\phi_+ + \phi_- = 0$. The scattering matrix of the wall is simply the reflection coefficient $S_{\text{wall}} = r_{\text{wall}}$ such that $\phi_- = r_{\text{wall}} \phi_+$, so $r_{\text{wall}} = -1$.}.
Hence, the resonance (constructive interference) condition in the cavity formed by the upper wall and the gapped stratification is
\begin{equation}
	\label{constructive_interference_condition}
	r'(\omega) \times r_{\text{wall}} = 1
\end{equation}
i.e. $\varphi_0'(\omega) + 2 k_1(\omega) \ell + \pi = 2 \pi n$ with $n \in \mathbb{Z}$.
This condition is usually satisfied only for discrete values of $\omega$ that are the frequencies of the surface modes.
Note that in general, the condition \eqref{constructive_interference_condition} can hold even when $\ell = 0$ (see Fig.~\ref{surface_states_simplified_model_uc_ratio} and discussion below for an example). 

We emphasize that the walls are not a necessary ingredient: the same mechanism can take place at an interface between two gapped stratifications. It is also not necessary that the gapped stratifications are spatially periodic: they can also be quasiperiodic (see Ref.~\cite{Levy2017} and references therein for examples in photonic crystals) or disordered.

\medskip

Surface states can also be obtained by propagating an initial condition that encodes the boundary condition into the crystal with the transfer matrix of the unit cell~\cite{Lee1981,Dwivedi2016} (see also Refs.~\cite{Hatsugai1993,Hatsugai1993b,Tauber2015,Kunst2019}, in particular for the relation with topological band theory). States that decay from the boundary correspond to surface states, while those with constant amplitude are bulk modes. This perspective is convenient for the numerical (or analytical) computation of the frequencies of surface states. 
To do so, it is more practical to start from the bulk and then match the boundary condition. First, consider the eigenvectors $\Psi$ of $T_{\text{uc}}(\omega)$ with eigenvalues $\lambda$ satisfying $|\lambda| > 1$, so that they decay in the crystal from an edge located on the right side of the system (provided that the appropriate boundary condition is satisfied).
Second, express the boundary condition $\phi^{\text{wall}} = 0$ as
$\mathcal{W}(\Psi^{\text{wall}}) \equiv \phi_+^{\text{wall}} + \phi_-^{\text{wall}} = 0$.
Hence, surface states correspond to the eigenvectors $\Psi$ of $T_{\text{uc}}$ with eigenvalues $|\lambda| > 1$ that satisfy $\mathcal{W}(\Psi) = 0$ when the crystal is terminated at the end of a unit cell.
Solving for the frequencies $\omega_{\text{ss}}$ such that both conditions are simultaneously satisfied gives the frequencies of the surface modes.
When the last unit cell is incomplete of modified, we have to first express the fields $\Psi^{\text{wall}}$ at the wall in terms of the fields $\Psi^{\text{end}}$ at the end of the crystal (i.e., of the last whole unit cell) by an appropriate transfer matrix $T_{\text{surf}}$ such that $\Psi^{\text{wall}} = T_{\text{surf}} \Psi^{\text{end}}$, and to ask that the eigenvector satisfies $\mathcal{W}(T_{\text{surf}} \Psi) = 0$ instead.

\begin{figure}
\includegraphics[width=2.8in]{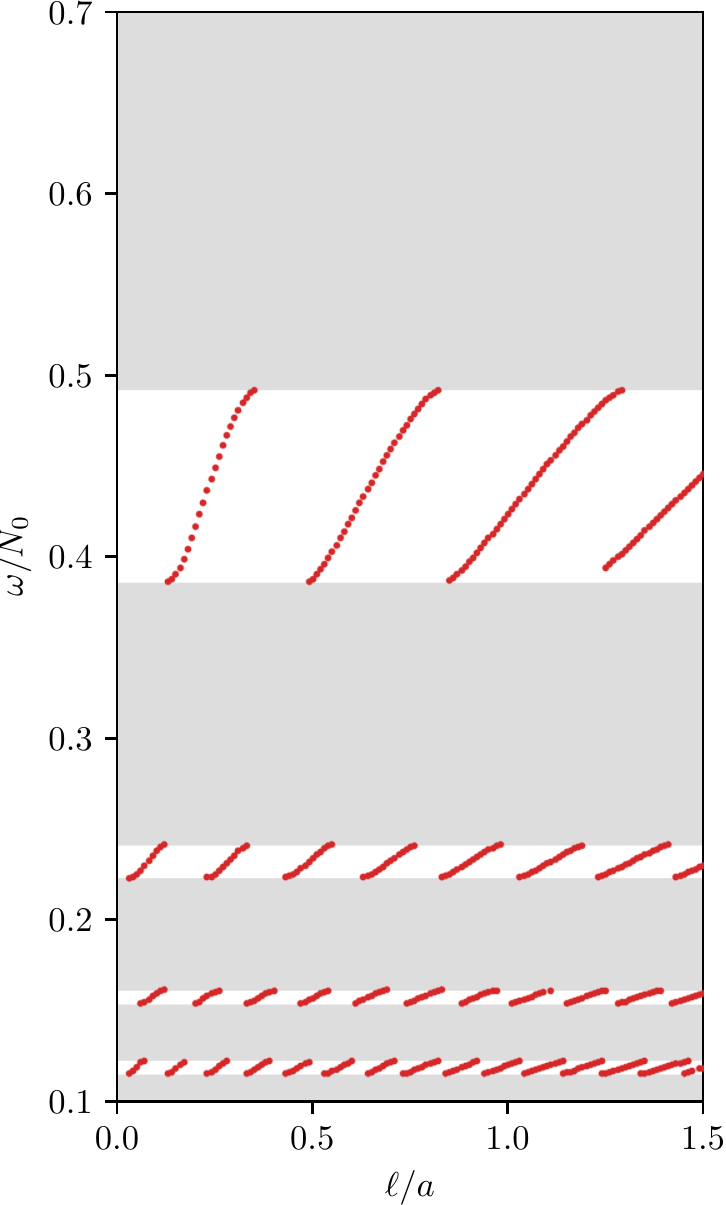}
\caption{\label{surface_states_simplified_model_end_layer} \textbf{Surface states of the simplified model.}
Surface states of the simplified model as a function of the excess thickness of the last layer $\ell$ (see schematic on Fig.~\ref{setup_surface_modes_scattering}), computed using the transfer matrix method~\cite{Lee1981,Dwivedi2016}.
Here, $\ell_1/a=\ell_2/a = 1/2$.
The bands are marked as grey regions, and the surface states as red curves.
We have only drawn the surface states at a single wall, where the boundary condition $\psi = 0$ is imposed.
Unless otherwise specified, the parameters are the same as in Fig.~\ref{surface_states_simplified_model_uc_ratio}.
}
\end{figure}

In Fig.~\ref{surface_states_simplified_model_uc_ratio}, we compute in this way the bulk band structure (bulk bands are drawn in grey) and the frequencies of the top surface states (in red) in a generalization of our simplified model, in which the two layers have different thicknesses $\ell_1$ and $\ell_2$ (with $a = \ell_1 + \ell_2$) and the additional layer is not present ($\ell = 0$; the effect of a non-zero $\ell$ is represented on Fig.~\ref{surface_states_simplified_model_end_layer}). 
When $\omega$ is in a bulk gap, the scattering matrix corresponding to the transfer matrix for $n$ unit cells $[T_{\text{uc}}(\omega)]^n$ is approximately diagonal (or block-diagonal if there are more than one channel) as the transmission coefficients on the off-diagonal decay exponentially with $n$. The reflection coefficients on the diagonal converge to the value of the infinite crystal as $n \to \infty$. In practice, a few unit cells are enough to obtain good approximate convergence.
In the inset of Fig.~\ref{surface_states_simplified_model_uc_ratio}, we plot the phase $\arg r'$ of the reflection coefficient at the surface for a fixed $\omega$ corresponding to the dashed blue line in Fig.~\ref{surface_states_simplified_model_uc_ratio} and $n=\num{20}$.
As can be seen on the figure, a surface state indeed occurs when $\arg r' = \pi$, which corresponds to the constructive interference condition $r'(\omega) \times r_{\text{wall}} = 1$ (with $r_{\text{wall}} = -1$) discussed above.
We also observe on Fig.~\ref{surface_states_simplified_model_uc_ratio} that continuously changing the parameter $\ell_2/[\ell_1+\ell_2]$ moves a surface state from one band to another: each surface state (in red) connects the band above its gap to the band below as the parameter is changed~\footnote{Expanding $k_z(\omega)$ and $\varphi_0'(\omega)$ in series around $\omega^0$ by writing $\omega = \omega^0 + \Delta \omega$, the condition of constructive interference \eqref{constructive_interference_condition} becomes $\alpha + \beta \Delta \omega = 2 \pi n$ (in which $\alpha$ and $\beta$ are coefficients obtained from the expansion).
Starting from a solution $\Delta \omega_{0}$ of this equation, we change the reflection phase of the crystal by $\Delta \varphi$ (independently of frequency for simplicity). This amounts to the change $\alpha \to \alpha + \Delta \varphi$. Hence, the frequency of the surface state becomes $\Delta \omega_{0} - \Delta \phi/\beta$.}. 
Formally, this is similar to a Thouless pump~\cite{Thouless1983}. 
Physically, it means that a two-dimensional system in which the reflection phase is continuously changed in a dimension orthogonal to the direction of periodicity (for instance by shifting the internal wave crystal linearly) will generically exhibit non-zero first Chern numbers; this has been demonstrated in optics in Ref.~\cite{Nakata2020} (see also Ref.~\cite{Poshakinskiy2015}, in which the additional dimension is a control parameter).


\medskip

We now comment on the relation between surface states and topology. 
In their simplest form, topological insulators are spatially periodic systems (crystals) in which (i) the band structure describing wave propagation exhibits gaps and (ii) certain topological properties (winding numbers and their generalizations) can be ascribed to the bulk Bloch bands. 
These topological states occur in various contexts including electrons in solids~\cite{Hasan2010}, cold atoms~\cite{Cooper2019}, optics~\cite{Ozawa2019,Lu2014}, acoustic and mechanics~\cite{Huber2016,Ma2019}, but also in biochemical networks~\cite{Murugan2017}, active matter~\cite{Shankar2020}, and geophysical fluids~\cite{Delplace2017,Perrot2019}.
In many cases (but not all), a relation called the bulk-boundary correspondence connects the topological properties of the bulk bands to the existence of surface states at an interface.
In continuum models, however, the bulk-boundary correspondence is often modified to account for subtleties not present in lattice models, and is therefore less easily applicable~\cite{Silaev2012,Souslov2019,Tauber2019,Tauber2020,Silveirinha2015,Silveirinha2016,Silveirinha2019}.

\begin{figure}
\includegraphics[width=3.2in]{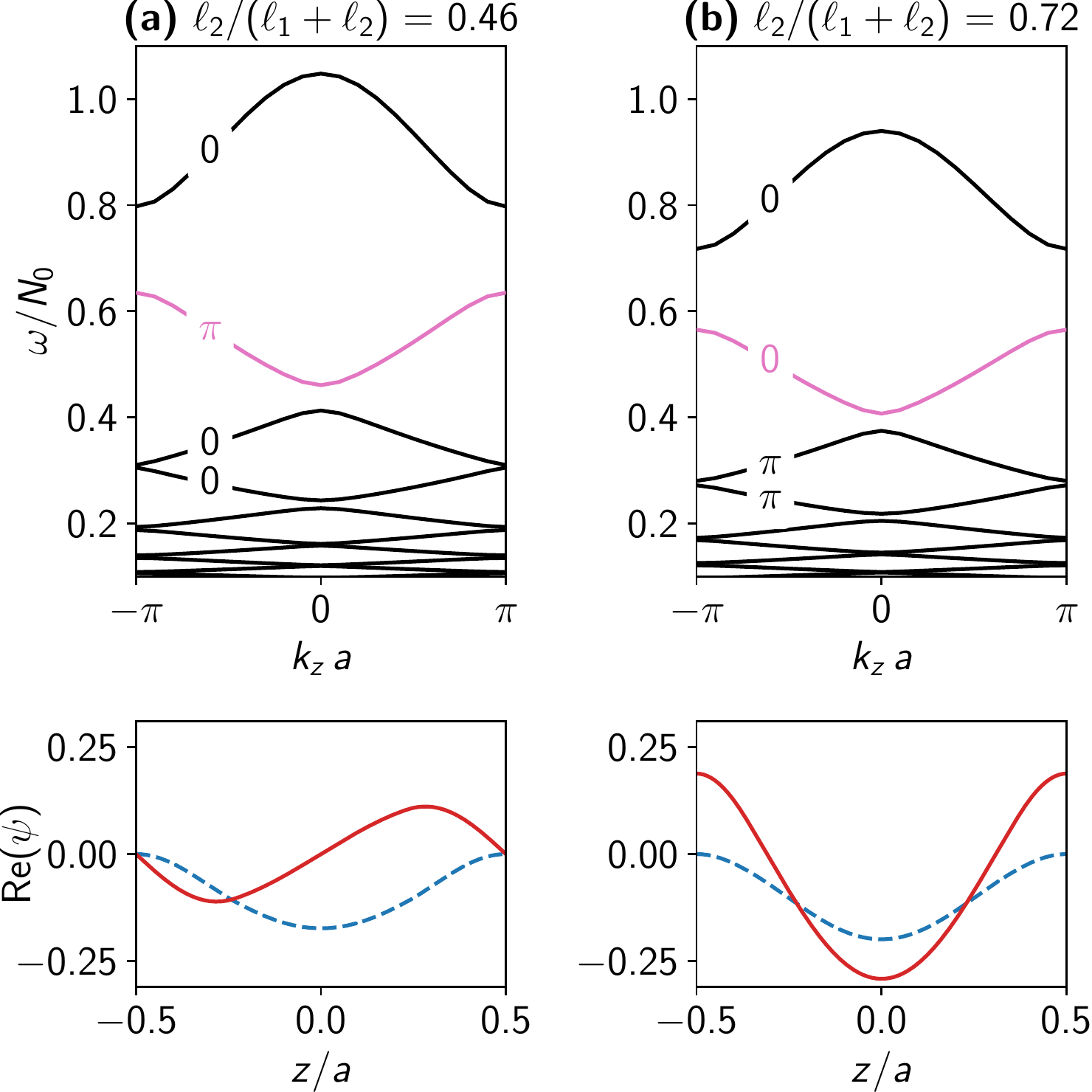}
\caption{\label{figure_zak_phases} \textbf{Zak phases in the simplified model.}
The band structures of the simplified model are computed with the spectral method described in Sec.~\ref{computation_bs_spectral} for two values of $\ell_2/[\ell_1 + \ell_2]$, respectively in panels (a) and (b).
We also compute the Zak phases of the four first (highest-frequency) bands using Eq.~\eqref{zak_phase}.
In the bottom panels, we show the (real part of) the Bloch wavefunction of the second highest-frequency band (highlighted in pink in the top panels), at $q=0$ (dashed blue lines) and $q=\pi/a$ (continuous red lines) in both cases. In (a), they have opposite symmetries with respect to inversion $z \to -z$, so the Zak phase is $\pi$. In (b), they have the same symmetry, so the Zak phase is $0$.
We have set $N_1/N_0 = \num{1.2}$, $N_2/N_0 = \num{0.8}$, $k_x/a = \num{3.0}$. 
}
\end{figure}

\begin{figure}
\includegraphics[width=2.0in]{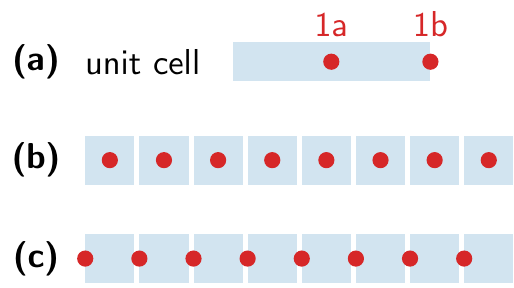}
\caption{\label{figure_wyckoff} \textbf{Topological phases and maximal Wyckoff positions.}
We show in (a) the two maximal Wyckoff positions 1a and 1b in a one-dimensional unit cell with spatial inversion symmetry. 
Panels (b) and (c) show a sketch of the two possible atomic limits, corresponding to Wannier centers (in red) pinned either at the Wyckoff position 1a (at the center of the unit cell) or at the Wyckoff position 1b (at the edge of the unit cell). 
(Both edges of the unit cell are equivalent; the spacing between the unit cells has been added for clarity and is not really present.)
}
\end{figure}

Our simplified model with alternating stratifications $N_1$ and $N_2$ is similar to continuum models of photonic crystals made of alternating dielectric slabs with permittivities $\epsilon_1$ and $\epsilon_2$. 
The surface states are similar to Tamm-Shockley states~\cite{Tamm1932,Shockley1939} that exist in electronic, optical, and other wave systems.
The topological nature of these surface states in optics and acoustics has been discussed in Refs.~\cite{Xiao2014,Levy2017,Henriques2020} (see also references therein). 
The continuum photonic crystal analysed in Refs.~\cite{Xiao2014,Levy2017,Henriques2020} can indeed be mapped~\cite{Henriques2020} to the Su-Schrieffer-Heeger (SSH) model, a standard example of symmetry-protected topological insulator~\cite{Hasan2010,Asboth2016} (which involves some subtleties \cite{Shapiro2021,Fuchs2021,Guzman2020}). The same applies to our models of internal waves in periodic stratifications. 
The relevant topological invariant is called the Zak phase $\phi_{\text{Zak}}$ (there is one Zak phase per band).
In general, the Zak phase can be defined as the integral of the Berry connection across the one-dimensional Brillouin zone~\cite{Zak1989}, and it can take any value. 
In a system with spatial inversion symmetry (here, $z \to -z$), the Zak phase is constrained to take values $0$ or $\pi$, and can be expressed as~\cite{Zak1989,Wang2019}
\begin{equation}
	\label{zak_phase}
	\frac{\phi_{\text{Zak}}}{\pi} = \frac{\xi_0 - \xi_\pi}{2} \, \text{mod $2$}
\end{equation}
in which $\xi_{q}$ is the inversion eigenvalue (or parity eigenvalue) of the band at wave-vector $q=0,\pi/a$, which is $\pm 1$ depending on whether the eigenvector $\psi(k)$ describing the band is symmetric or antisymmetric (i.e., even or odd under inversion). The two possible values $\phi_{\text{Zak}}=0,\pi$ can be considered as topologically inequivalent. 
To see why, note that the Zak phase gives, qualitatively, the average position of the wave in the unit cell.
Formally, this quantity is called the Wannier center of the band (see Refs.~\cite{Zak1989,Bradlyn2017,Cano2021,Wieder2021} for details).
It turns out that when the system is invariant under spatial symmetries, the Wannier center must be located at special points in the unit cell called (maximal) Wyckoff positions. 
In a system with inversion symmetry, the Wannier center can only be located at the center or at the edge of the unit cell (the Wyckoff positions 1a and 1b, see Fig.~\ref{figure_wyckoff}), that correspond respectively to $\phi_{\text{Zak}}=0$ and $\phi_{\text{Zak}}=\pi$. 
In Fig.~\ref{figure_zak_phases}, we show the band structure of the simplified model labelled with their Zak phases calculated from Eq.~\eqref{zak_phase}. A topological transition occurs between the cases (a) and (b), when the bulk band gap around $\omega/N_0 \sim \num{0.25}$ closes at $\ell_2/(\ell_1+\ell_2) \sim \num{0.6}$ (see Fig.~\ref{figure_zak_phases}).
In the framework of topological quantum chemistry~\cite{Bradlyn2017,Cano2021,Wieder2021}, the bands with $\phi_{\text{Zak}}=\pi$ (equivalently, with Wannier centers at Wyckoff position 1b) are in an obstructed atomic limit. In contrast, those with $\phi_{\text{Zak}}=0$ (Wyckoff position 1a) are in a trivial atomic limit.
(For more details, see Fig.~1 of Ref.~\cite{Wieder2021}, Sec.~III of Ref.~\cite{Benalcazar2019}, and SI Sec.~IV of Ref.~\cite{Bradlyn2017}.)

We can also ask to what extent does it matter in practice that a surface state is topological, besides the fundamental interest of the underlying mathematical structure.
In certain contexts (such as the quantum Hall effect), the topological origin of edge states confers them a robustness against defects, disorder, and changes in the system's parameter. Here, however, the existence and properties of surface states depends strongly on how the system is terminated (this can already be seen in Fig.~\ref{surface_states_simplified_model_uc_ratio}).

\bibliography{bibliography.bib}

\end{document}